\documentclass[preprint, 5p, twocolumn]{elsarticle}

\usepackage{lineno,hyperref}
\usepackage{lineno}
\usepackage{graphicx}
\usepackage{subcaption}
\usepackage{amsmath}
\usepackage{verbatim}
\usepackage{amssymb}
\usepackage{color}
\usepackage{xcolor}

\usepackage{mathrsfs}
\modulolinenumbers[5]

\journal{Journal of \LaTeX\ Templates}

\begin{document}

\begin{frontmatter}

\title{K-field kinks in two-dimensional dilaton gravity}


\author[mysecondaryaddress]{Yuan Zhong\corref{mycorrespondingauthor}}
\cortext[mycorrespondingauthor]{Corresponding author}
\ead{zhongy@mail.xjtu.edu.cn}
\author{Fei-Yu Li}
\author{Xu-Dong Liu}
\address{School of Physics, Xi'an Jiaotong University, \\
No. 28, West Xianning Road, Xi'an 710049, People's Republic of China}

\begin{abstract}
In this work, kinks with non-canonical kinetic energy terms are studied in a type of two-dimensional dilaton gravity model. The linear stability issue is generally discussed for arbitrary static solutions, and the stability criteria are obtained. As an explicit example, a model with cuscuton term is studied. After rewriting the equations of motion into simpler first-order formalism and choosing a polynomial superpotential, an exact self-gravitating kink solution is obtained. The impacts of the cuscuton term are discussed.
\end{abstract}


\end{frontmatter}



\section{Introduction}

Comparing to their higher-dimensional counterparts, gravitational models in two dimensions usually have simpler dynamics. For this reason, two-dimensional (2D) gravity models have been applied in the study of difficult issues like quantum gravity~\cite{Henneaux1985,Alwis1992}, gravitational collapse~\cite{VazWitten1994,VazWitten1996}, black hole evaporation~\cite{CallanGiddingsHarveyStrominger1992,BilalCallan1993,RussoSusskindThorlacius1992,RussoSusskindThorlacius1992a,RussoSusskindThorlacius1993},  see~\cite{Brown1988,Thorlacius1995,NojiriOdintsov2001d,GrumillerKummerVassilevich2002} for comprehensive reviews on early works. Recently, the studies of the Sachdev-Ye-Kitaev (SYK) model \cite{SachdevYe1993,Kitaev2015} also lead to a resurgence of interest in 2D gravity~\cite{AlmheiriPolchinski2015,MaldacenaStanfordYang2016,MaldacenaStanford2016,Jensen2016}, see~\cite{Rosenhaus2019,Sarosi2018,Trunin2021} for pedagogical introductions.

Because the Einstein tensor is zero identically in two dimensions, one must extend the Einstein-Hilbert action in order to describe 2D gravity. An economical way of extending Einstein's gravity is to introduce a dilaton field $\varphi$, which couples with metric non-minimally. The simplest example is the Jackiw-Teitelboim (JT) gravity, whose action reads~\cite{Jackiw1985,Teitelboim1983}
\begin{eqnarray}
S_{J T}=\frac{1}{\kappa} \int d^{2} x  \sqrt{-g}[\varphi(R+\Lambda)+\kappa \mathcal L_{M}],
\end{eqnarray}
where $\kappa$ and $\Lambda$ are the gravitational coupling and the cosmological constant, respectively. $\mathcal L_{M}$ represents the Lagrangian density of additional matter fields. Note that the dilaton in JT gravity is merely a Lagrangian multiplier, which has no dynamics. Consequently, the energy momentum tensor is not covariantly conserved. A natural generalization of the JT gravity which yields conserved energy momentum is proposed by Mann, Morsink, Sikkema and Steele (MMSS)~\cite{MannMorsinkSikkemaSteele1991}, who added a kinetic term to the dilaton field. Omitting the cosmological constant, the action of MMSS model takes the following form
\begin{eqnarray}
\label{MMSSgra}
S_{\textrm{MMSS}}=\frac{1}{\kappa}\int{d^2x}\sqrt{-g}\left[ -\frac{1}{2}(\nabla\varphi)^2  +\varphi  R  +\kappa \mathcal L_{M}\right].
\end{eqnarray}
This action can also be understood as the $D\to2$ limit of general relativity~\cite{MannRoss1993}.
Many issues of 2D gravity have been studied previously under this model, including N-body motion~\cite{OhtaMann1996,MannOhta1997,MannOhta1997a,MannPotvinRaiteri2000,BurnellMannOhta2003},  black hole chemistry~\cite{FrassinoMannMureika2015}, entropic gravity~\cite{MannMureika2011}, and the creation of primordial black holes~\cite{TzikasNicoliniMureikaCarr2018}.  On the other hand, regular solutions generated by nonlinear matter fields are rarely discussed.

In this work, we consider a classical solution with localized energy density called kink. Kink has been applied in many different branches of modern physics, ranging from condensed matter physics to cosmology~\cite{Vachaspati2006}. For example, in some modern theories of extra dimensions, our world is assumed to be a kink interpolating between two AdS$_5$ spaces. Such self-gravitating kink solutions are also known as thick branes~\cite{SkenderisTownsend1999,DeWolfeFreedmanGubserKarch2000,Gremm2000}, which can be regarded as the regular extensions of the Randall-Sundrum thin brane world models~\cite{RandallSundrum1999,RandallSundrum1999a,GoldbergerWise1999,GoldbergerWise1999a}, see~\cite{DzhunushalievFolomeevMinamitsuji2010,Liu2018} for reviews on thick branes.   In parallel to the study of 2D black holes, it is natural to consider 2D thick brane models to see if interesting results would be obtained.

In fact, an exact 2D self-gravitating kink solution had been obtained early in 1995 by St\"otzel~\cite{Stoetzel1995}, who considered the MMSS model with a sine-Gordon scalar. This solution can be regarded as a 2D thick brane solution, since the kink connects two AdS$_2$ spaces. Interestingly, in a recent work~\cite{Zhong2021}, one of the present authors showed that the method  used by St\"otzel in deriving his solution is nothing but the superpotential method, which has been widely applied in the study of 5D thick branes.  With this method, many other 2D self-gravitating kink solutions can be easily constructed. Moreover, the linear stability of these solutions were analyzed by using the factorization method~\cite{Zhong2021}. These results indicate that the MMSS gravity is a suitable platform for studying 2D thick brane solutions.

The scalar fields considered in Refs.~\cite{Stoetzel1995,Zhong2021} have canonical dynamics, in other words, the Lagrangian density of the scalar is simply $\mathcal{L}_{M}=X-V(\phi)$, where $X=-\frac{1}{2}g^{\mu\nu} \nabla_{\mu} \phi \nabla_{\nu} \phi$ is the kinetic term of the scalar field $\phi$.
It would be interesting to study a model with non-canonical scalar field, for which $\mathcal{L}_{M}=\mathcal{L}(X,\phi)$. Scalar field of this type is called K-field, which was first introduced in the study of cosmology~\cite{Armendariz-PiconDamourMukhanov1999,GarrigaMukhanov1999}, then was applied to construct topological defect solutions of various dimensions~\cite{Babichev2006,BazeiaLosanoMenezesOliveira2007,AdamSanchez-GuillenWereszczynski2007,Babichev2008}, including some 5D thick brane solutions~\cite{AdamGrandiSanchez-GuillenWereszczynski2008,BazeiaGomesLosanoMenezes2009,LiuZhongYang2010,ZhongLiuZhao2014a,ZhongLiu2013}.

The aim of the present work is twofold. Firstly, to establish stability criteria for general static K-field kinks in the MMSS gravity. We will show in the next section that if the Lagrangian density of the K-field satisfies two conditions, the corresponding solutions will always be stable. Secondly, to construct an exact K-field kink solution in the MMSS gravity. With the stability criteria in mind, we then consider an explicit K-field model, i.e., a model with cuscuton term in Sec.~\ref{SecThree}. Despite the appearance of a non-canonical cuscuton term, the dynamical equations can still be rewritten as simple first-order formalism, from which exact self-gravitating kink solutions can be easily constructed. Our conclusion will be given in Sec.~\ref{SecConc}.

\section{Model and stability analysis}
\label{SecTwo}
Our model takes the following action
\begin{equation}\label{1}
S=\frac{1}{\kappa} \int d^{2} x \sqrt{-g}\left[-\frac{1}{2} \nabla^{\mu} \varphi \nabla_{\mu} \varphi+\varphi R+\kappa \mathcal{L}(\phi,X)\right],
\end{equation}
which, after variations, leads to three dynamical equations, namely, the Einstein equation
\begin{eqnarray}
\label{STeq1}
&&\nabla_{\mu} \varphi \nabla_{\nu} \varphi-\frac{1}{2} g_{\mu \nu}\left(\nabla^{\lambda} \varphi \nabla_{\lambda} \varphi+4 \nabla_{\lambda} \nabla^{\lambda} \varphi\right)\nonumber\\
&+&2 \nabla_{\mu} \nabla_{\nu} \varphi=-\kappa T_{\mu \nu},
\end{eqnarray}
the dilaton equation
\begin{equation}
\label{STeq2}
\nabla^{\lambda} \nabla_{\lambda} \varphi+R=0,
\end{equation}
and the scalar equation
\begin{eqnarray}
\label{STeq3}
\mathcal{L}_{X} \nabla_{\lambda} \nabla^{\lambda} \phi+\nabla_{\lambda} \mathcal{L}_{X}\nabla^{\lambda} \phi+\mathcal{L}_{\phi}=0.
\end{eqnarray}
The energy-momentum tensor in Eq.~\eqref{STeq1} is defined by
\begin{equation}
T_{\mu \nu}=g_{\mu \nu}\mathcal{L}+ \mathcal{L}_X \nabla _{\mu}\phi \nabla _{\nu}\phi.
\end{equation}
In this work, subscriptions $X,\phi,\cdots$ are used to represent the derivatives of $\mathcal{L}$, thus, $\mathcal{L}_{X}\equiv \frac{\partial\mathcal{L}}{ \partial{X}}$ and $\mathcal{L}_{\phi}\equiv \frac{\partial\mathcal{L}}{ \partial{\phi}}$.

As in Refs.~\cite{Stoetzel1995,Zhong2021}, we will look for self-gravitating kink solutions with the following metric:
\begin{equation}
\label{metricXCord}
  ds^2=-e^{2A(x)}dt^2+dx^2,
\end{equation}
where $A(x)$ is the warp factor. Under this coordinates, the dilaton equation \eqref{STeq2} reads
\begin{equation}
\partial_x^2 \varphi+\partial_x A \partial_x \varphi=2 \partial_x^2 A+2 (\partial_x A)^{2},
\end{equation}
which has a simple solution~\cite{Stoetzel1995,Zhong2021}, namely, $\varphi(x)=2A(x)$. After replacing $\varphi(x)$ by $A(x)$, the Einstein equations become
\begin{eqnarray}
\label{eqEin1}
-4 \partial_x^2 A&=&\kappa  \mathcal{L}_X (\partial_x\phi)^2, \\
\label{eqEin2}
4 \partial_x^2 A +2 \left(\partial_x A\right)^2&=&\kappa  \mathcal{L}.
\end{eqnarray}
The scalar equation takes the following form
\begin{equation}
\mathcal{L}_X \left(\partial_x A \partial_x \phi+\partial_x^2 \phi\right)+\partial_x \mathcal{L}_{{X }} \partial_x \phi+\mathcal{L}_{\phi }=0,
\end{equation}
which can also be derived from the Einstein equations, and therefore, is not independent from the later. Therefore, to find a solution of the system, we only need to solve Eqs.~\eqref{eqEin1} and \eqref{eqEin2}.

Before considering explicit K-field models, let us first conduct a general analysis on the stability of arbitrary static kink solutions under small linear perturbations. It is convenient to introduce a new spatial coordinate
\begin{equation}
r\equiv\int e^{-A(x)}dx,
\end{equation}
with which the line element becomes conformally flat
\begin{equation}
  ds^2=e^{2A(r)}(-dt^2+dr^2).
\end{equation}
For simplicity, we will use overdots and primes to denote the derivatives with respect to $t$ and $r$, respectively.

In the $r$-coordinates, the dilaton equation takes a simpler form
\begin{equation}
\varphi''=2 A''.
\end{equation}
As in the $x$-coordinates, we will omit the integral constants and take the solution of the dilaton equation as
\begin{equation}
\label{EqDilatonR}
\varphi(r)=2A(r).
\end{equation}
Then, the Einstein equations and the scalar equation become
\begin{eqnarray}
\label{EqEin1R}
4 A''-2A'^2&=&\kappa e^{2 A} \mathcal{L},\\
\label{EqEin2R}
2 A''- 2A'^2 &=&  \kappa e^{2 A} \mathcal{L}_X X,
\end{eqnarray}
and
\begin{equation}
\label{EqScaR}
\mathcal{L}_X \phi ''+\phi ' \mathcal{L}_X'+e^{2 A}  \mathcal{L}_{\phi }=0,
\end{equation}
respectively.

Following Ref.~\cite{Zhong2021}, let us now consider small perturbations $\{\delta\varphi(r,t), \delta\phi(r,t), \delta g_{\mu\nu}(r,t)\}$ around an arbitrary static solution of Eqs.~\eqref{EqDilatonR}-\eqref{EqScaR}, say, $\{\varphi(r), \phi(r), g_{\mu\nu}(r)\}$. It is convenient to rewrite the perturbation of the metric as follows
\begin{eqnarray}
\delta g_{\mu\nu}(r,t)&\equiv& e^{2A(r)} h_{\mu\nu}(r,t)\nonumber\\
&=&e^{2A(r)} \left(
\begin{array}{cc}
 h_{00}(r,t) & \Phi (r,t) \\
 \Phi (r,t) & h_{rr}(r,t) \\
\end{array}
\right).
\end{eqnarray}

To the first order, the perturbation of the metric inverse is given by
\begin{equation}
\delta g^{\mu \nu}=-e^{-2A} h^{\mu \nu}.
\end{equation}
Note that the indices of $h$ are always raised or lowered with $\eta_{\mu\nu}$, thus,
\begin{equation}
h^{\mu \nu}\equiv \eta^{\mu\rho}\eta^{\nu\sigma}h_{\rho\sigma}=\left(
\begin{array}{cc}
 h_{00} & -\Phi \\
 -\Phi & h_{rr} \\
\end{array}
\right).
\end{equation}
As  in Ref.~\cite{Zhong2021}, we define a new variable $\Xi \equiv 2 \dot{\Phi}-h_{00}^{\prime}$, and conduct our calculations in the dilaton gauge, where $\delta \varphi=0$.

The linearization of the Einstein equations \eqref{STeq1} leads to two independent perturbation equations, namely,
the $(0,1)$ component:
\begin{equation}
\label{eqPertOne}
2 A' {h}_{rr}=\kappa  \mathcal{L}_X \phi ' {\delta \phi } ,
\end{equation}
and the $(1,1)$ component:
\begin{eqnarray}
\label{eqPertTwo}
\Xi &=&2 \gamma \left(A'-\frac{A''}{A'}\right) \frac{ \delta \phi ' }{\phi '}
+\gamma \left[2\left(A'-\frac{A''}{A'}\right)^2 \right.\nonumber\\
&-&\left. \left(A'  - \frac{A'' }{ A'} \right)\frac{X'}{X}\right] \frac{  \delta \phi  }{\phi '},
\end{eqnarray}
where we have defined
\begin{equation}
\gamma\equiv 1+2\frac{\mathcal{L}_{XX}X}{\mathcal{L}_{X}}.
\end{equation}
The $(0,0)$ component is also nontrivial, but after substituting the background equations it reduces to Eq.~\eqref{eqPertOne}.

Another independent perturbation equation can be obtained from the linearization of the scalar equation \eqref{STeq3}, which, after substituting Eqs. \eqref{eqPertOne} and \eqref{eqPertTwo}, takes the following form:
\begin{eqnarray}
&&-\ddot{\delta \phi }
+\gamma  \delta \phi ''
+\gamma  \left(\frac{\gamma '}{\gamma}
+\frac{  \mathcal{L}_{X}'}{{\mathcal{L}_{X}}}\right)\delta \phi ' \nonumber\\
		&& +\gamma
	 \left[
	 	A'^2
		-3 A''
		+2\frac{ A'''}{A'}
		- 2 \left(\frac{A''}{A'}\right)^2\right.\nonumber\\
		&&\left.
		-\frac12 \frac{\mathcal{L}_X ' }{ \mathcal{L}_X } \frac{X'}{ X}
		-\frac12\frac{X''}{ X}
		+\frac{1}{4} \left(\frac{X'}{X}\right)^2
	\right]\delta \phi \nonumber\\
		&& +\gamma '
  \left(
	\frac{A''}{A'}
	-\frac12\frac{X'}{ X}	-A' \right)\delta \phi =0.
\end{eqnarray}
To proceed, we introduce a new variable
\begin{equation}
G(r,t) \equiv  \mathcal{L}_X^ {1/2}\gamma^{1/4 }  \delta\phi(r,t),
\end{equation}
and conduct another spatial coordinate transformation $r\to y$, such that
$dy/dr= \gamma^{-1/2}$. Finally, we find that $G(y,t)$ satisfies a compact wave equation:
\begin{equation}
-\partial_{t}^2 G+\partial_{y}^2 G- V_{\text{eff}}(y)G =0,
\end{equation}
where
\begin{equation}
V_{\text{eff}}(y)\equiv \frac{\partial_y^2 f}{f},
\end{equation}
and
\begin{equation}
 f(y)\equiv \mathcal{L}_X^{1/2} \gamma ^{1/4} \frac{\partial_y\phi}{\partial_y A}.
\end{equation}
After conducting the modes expansion
\begin{equation}
G(y,t)=\sum_n e^{i\omega_n t}\psi_n(y),
\end{equation}
one immediately obtains a Schr\"odinger-like equation of $\psi_n(y)$:
\begin{equation}
\label{eqSch}
\hat{H} \psi_n\equiv -\frac{d^2\psi_n}{dy^2}+V_{\text{eff}}(y)\psi_n=\omega_n^2 \psi_n.
\end{equation}
The particular form of $V_{\text{eff}}(y)$ enables us to factorize the Hamiltonian operator into the following form:
\begin{equation}
\label{eqFact}
\hat H =\hat{ \mathcal{A}}^\dagger \hat{ \mathcal{A}},
\end{equation}
where
\begin{equation}
\hat{\mathcal{A}}=-\frac{d}{d y}+\frac{\partial_y f}{f}, \quad \hat{\mathcal{A}}^{\dagger}=\frac{d}{d y}+\frac{\partial_y f}{f}.
 \end{equation}

The factorization of the Hamiltonian operator in the above form ensures that the eigenvalues of Eq.~\eqref{eqSch} are greater or equal to zero, so the operator is non-negative~\cite{InfeldHull1951}. Therefore, any static solution is stable against linear perturbations, if the scalar Lagrangian satisfies the following conditions:
\begin{equation}
\label{stabilityCond}
\mathcal{L}_X>0,\quad \gamma>0.
 \end{equation}

The ground state of Eq.~\eqref{eqSch} is the zero mode $\psi_0(y)$, which has zero eigenvalue and satisfies $\mathcal{A}^{\dagger}\psi_0(y)=0$. It is not difficult to show that the zero mode takes the following expression:
\begin{equation}
\label{ExpreZeroMode}
\psi_0(y)\propto f=\mathcal{L}_X^{1/2} \gamma ^{1/4} \frac{\partial_y\phi}{\partial_y A}.
\end{equation}

It is interesting to see that the stability criteria take the same form as those obtained for 5D thick K-branes in  Einstein gravity~\cite{ZhongLiu2013}. This result can be regard as another evidence of the conjecture of Ref.~\cite{MannRoss1993}, i.e., ``the theory based on action \eqref{MMSSgra} may be said to be the closest thing there is to general relativity in two dimensions".

The discussion in this section is rather general. In next section, we will solve a specific K-field model under the $x$-coordinates.

\section{A model with cuscuton term}
\label{SecThree}

\begin{figure*}[!ht]
\centering
\includegraphics[width=1\textwidth]{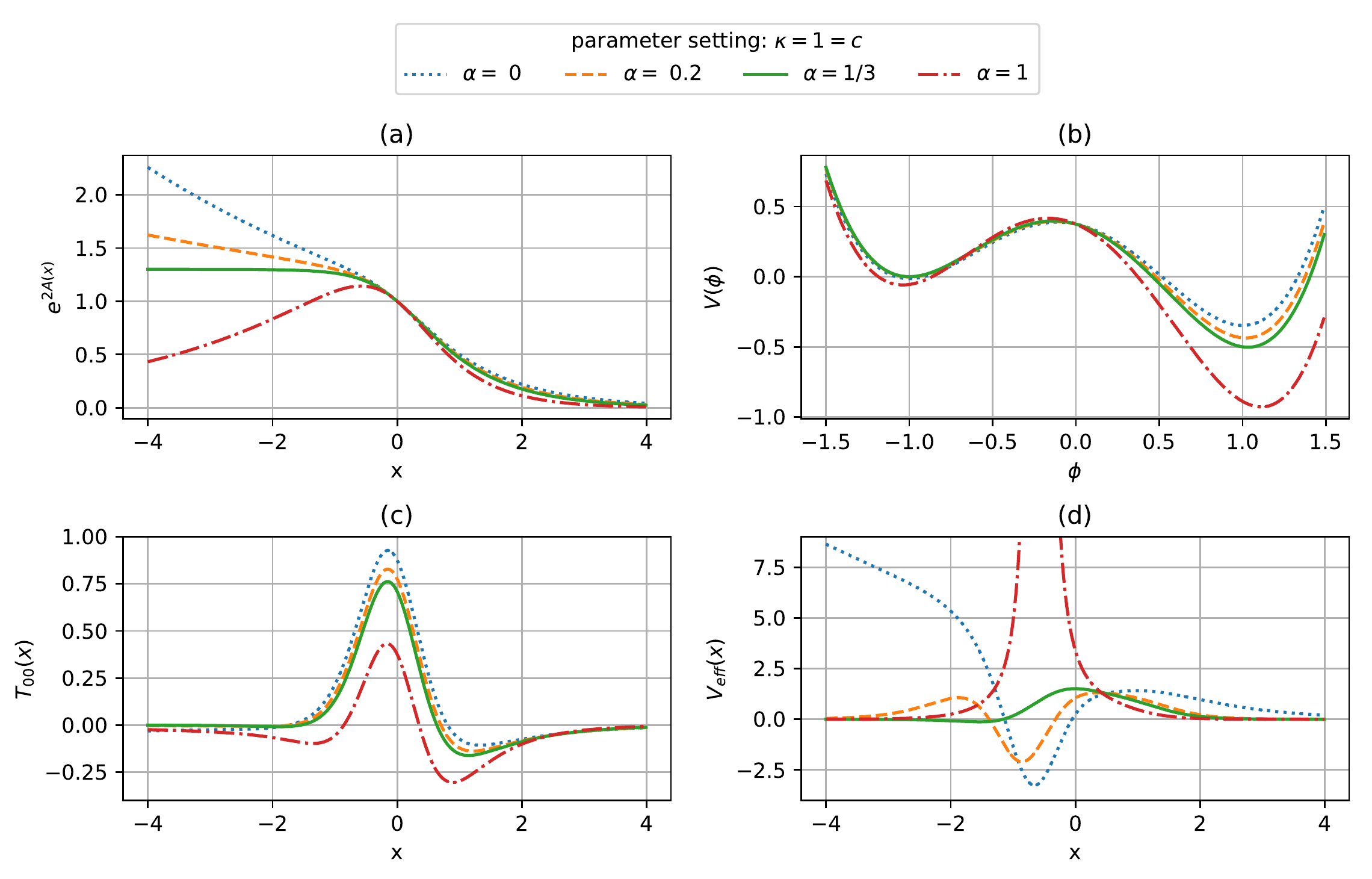}
\caption{Plots of (a) the warp factor $e^{2A(x)}$, (b) the scalar potential $V(\phi)$, (c) the energy density $T_{00}$, and (d) the effective potential $V_{\text{eff}}(x)$. The parameters are chosen as $\kappa=1=c$, and $\alpha=0, 0.2, 1/3, 1$. }
\label{Fig}
\end{figure*}
Cuscuton is a special subclass of K-fields which has the following Lagrangian density
\begin{equation}
\mathcal {L}=-\alpha\sqrt{-2X}-V(\phi),
\end{equation}
where $\alpha$ is a positive constant.
It was first introduced in Refs.~\cite{AfshordiChungGeshnizjani2007,AfshordiChungDoranGeshnizjani2007} as a new dark energy model. Unlike many other dark energy candidates, cuscuton is protected against quantum correction and has no dynamical degrees of freedom~\cite{AfshordiChungGeshnizjani2007,AfshordiChungDoranGeshnizjani2007,GomesGuariento2017}.

In addition to dark energy, cuscuton can also be used to construct regular models of bouncing cosmology~\cite{BoruahKimRoubenGeshnizjani2018,QuintinYoshida2020,KimGeshnizjani2021}, to reconcile the power-law inflation model with CMB data~\cite{ItoIyonagaKimSoda2019}, to accelerate universe with a stable extra dimension~\cite{ItoSakakiharaSoda2019}, and so on.

The Lagrangian we are going to study here, extends the cuscuton Lagrangian by including the standard kinetic term
\begin{equation}
\label{CuscLag}
\mathcal {L}=X-\alpha\sqrt{-2X}-V(\phi),
\end{equation}
for which
 \begin{equation}
 \label{eqLXGamma}
\mathcal {L}_X=1+ \alpha (-2X)^{-1/2}=\gamma^{-1}.
\end{equation}
 This model has been studied previously in the context of 5D thick brane model~\cite{BazeiaBritoCosta2013,AndradeMarquesMenezes2019,BazeiaFerreiraMarques2021}, where analytical thick brane solutions were obtained. Hopefully, similar stable self-gravitating kink solutions might be obtained in the present 2D model.

The Einstein equations \eqref{eqEin1}-\eqref{eqEin2} corresponding to Lagrangian \eqref{CuscLag} take the following form
 \begin{eqnarray}
\label{eqCusEin1}
-4 \partial_x^2 A&=& \kappa( \partial_x\phi)^2+\kappa \alpha\partial_x\phi, \\
\label{eqCusEin2}
4 \partial_x^2 A +2 \left(\partial_x A\right)^2&=&\kappa  [-\frac12( \partial_x\phi)^2-\alpha\partial_x\phi-V ].
\end{eqnarray}
Here we have used the fact that  a static kink  $\phi(x)$ is a monotonically increasing function of $x$, which means $\partial_x\phi\geq0$ for $x\in(-\infty, +\infty)$, and therefore $\sqrt{-2X}=|\partial_x\phi|=\partial_x\phi$. For the same reason, one can conclude from Eq.~\eqref{eqLXGamma} that the stability criteria $\mathcal {L}_X>0$ and $\gamma>0$ are always satisfied for arbitrary static kink solutions, provided $\alpha\geq0$.

Following Refs.~\cite{BazeiaBritoCosta2013,AndradeMarquesMenezes2019,BazeiaFerreiraMarques2021},  we apply the superpotential method to construct exact kink solutions of Eqs.~\eqref{eqCusEin1}-\eqref{eqCusEin2}. The essence of this method is to rewrite the dynamical equations into some first-order ones, i.e., the first-order formalism. To do this, we first introduce the superpotential function $W(\phi)$ via the following assumption
\begin{equation}
\label{EqFirsForm1}
\partial_x\phi=\frac{d W}{d \phi}.
\end{equation}
Then, from Eqs.~\eqref{eqCusEin1} and \eqref{eqCusEin2} one immediately obtains
\begin{equation}
\label{EqFirsForm2}
\partial_x A=-\frac14\kappa(W+\alpha \phi),
\end{equation}
and
\begin{equation}
\label{EqFirsForm3}
V=\frac{1}{2} W_\phi^2-\frac{1}{8} \kappa  (W+\alpha  \phi)^2,
\end{equation}
respectively. With the first-order formalism \eqref{EqFirsForm1}-\eqref{EqFirsForm3}, we are now ready to construct kink solutions by taking a suitable superpotential.

Here, we consider the same superpotential in Ref.~\cite{Zhong2021}
\begin{equation}
W=c+\phi-\frac{1}{3}\phi^3,
\end{equation}
where $c$ is a constant parameter. This superpotential was first introduced in some 5D thick brane models~\cite{EtoSakai2003,TakamizuMaeda2006,BazeiaMenezesRocha2014}, where interesting asymmetric thick brane solutions can be obtained if $c\neq0$. In our 2D case, the corresponding solution reads
\begin{eqnarray}
\label{solKink}
\phi(x)&=&\tanh(x),\\
\varphi(x)&=&2A(x),\\
\label{SolWarp}
A(x)&=&\frac{1}{24} \kappa  \left[-6 c x+\text{sech}^2(x)\right. \nonumber\\
&-&\left.2 (3 \alpha +2) \log (\cosh (x))-1\right],\\
V(\phi)&=&-\frac{1}{72} \kappa  \left(-3 c+\phi ^3-3 \phi -3 \alpha  \phi \right)^2\nonumber\\
&+&\frac{1}{2} \left(\phi ^2-1\right)^2.
\end{eqnarray}
When $\alpha=0$, this solution reduces to the one of Ref.~\cite{Zhong2021}, where the author has discussed the impacts of parameter $c$ by fixing $\kappa=1$. It would be interesting to see how the parameter $\alpha$ impacts the behavior of the solution.

Let us start by noticing that the cuscuton term modifies the asymptotic behavior of the warp factor:
\begin{equation}
\label{eqAsymA}
\lim_{x\to\pm\infty}\partial_x A= -\frac{1}{4} \kappa  \left(\frac{2}{3}+\alpha \pm c \right).
\end{equation}
Obviously, if $c=0$ the warp factor symmetrically connects two asymptotic AdS$_2$ spaces, whose radiuses can be tuned by the value of $\alpha$.

We are more interested in cases with $c\neq0$. Without loss of generality, we assume $c>0$, and simply take $c=1$ in what follows. From Eq.~\eqref{eqAsymA}, we see that there exists an interesting critical case, i.e., $\alpha=\alpha_{\textrm{cr}}\equiv c-\frac23=\frac13$, when the warp factor glues, asymptotically, a Minkowski space on the left side with an AdS space on the other side. This particular case is important for the study of brane collisions~\cite{TakamizuMaeda2006,OmotaniSaffinLouko2011}. While, if $\alpha>\alpha_{\textrm{cr}}$, the left and right sides approach to two different AdS$_2$ spaces. Finally, if $0\leq\alpha<\alpha_{\textrm{cr}}$, the left side of the warp factor diverges. These behaviors can be seen in Fig.~\ref{Fig}-(a), where we plotted  $e^{2A(x)}$ for $\kappa=1=c$ and $\alpha=0, 0.2, 1/3, 1$.

The cuscuton term also impacts the shape of $V(\phi)$, as can be seen in Fig.~\ref{Fig}-(b). Especially, we see that the positions of local minima varies with $\alpha$. This can be verified by numerical calculations. For example, the minimum at $\phi=1$ moves to $\phi \approx 1.0185, 1.0321, 1.1123$ for $\alpha= 0.2, 1/3, 1$, respectively. Therefore, when $\alpha\neq 0$ the kink solution of Eq.~\eqref{solKink} no longer connects two minima of the scalar potential.

The cuscuton term may also lead to the creation of inner structure (or milti-peak structure) of the energy density $T_{00}$, as shown in Ref.~\cite{BazeiaFerreiraMarques2021}.  In the present model, however, we did not observe this phenomenon within our parameter range, see Fig.~\ref{Fig}-(c), where $T_{00}$ only has a single peak.

Before ending this section, let us comment that the shape of the effective potential $V_{\text{eff}}$ will be significantly changed even for a small $\alpha$. To see this, we first transform $V_{\text{eff}}(y)$  back to the $x$-coordinates:
\begin{equation}
\label{eqVeffx}
V_{\text{eff}}(x)=f^{-1}  e^{A}\gamma^{1/2} \partial_x\left(e^{A}\gamma^{1/2} \partial_x f \right),
\end{equation}
with $f(x)=\mathcal{L}_X^{1/2} \gamma ^{1/4} \frac{\partial_x\phi}{\partial_x A}$.
Here we have used the following relation
\begin{equation}
\frac{dy}{dx}=\frac{dy}{dr}\frac{dr}{dx}=\gamma^{-1/2}e^{-A}.
\end{equation}
The analytical expression of $V_{\text{eff}}(x)$ can be obtained by simply inserting our solution into Eq.~\eqref{eqVeffx}.  Instead of  displaying a lengthy expression, we plot $V_{\text{eff}}(x)$ in Fig.~\ref{Fig}-(d), from which we clearly see that the cuscuton term mainly impacts the following two properties of $V_{\text{eff}}$:
\begin{enumerate}
\item The asymptotic behavior at $x\to -\infty$, where  $V_{\text{eff}}$ is divergent for $\alpha=0$, but becomes convergent as $\alpha>0$.
\item The depth of the center well, which decreases as $\alpha$ increases, until $\alpha>\alpha_{\textrm{cr}}$, when the well becomes an infinite barrier.
\end{enumerate}

The potential well existing in cases with $0\leq\alpha<\alpha_{\textrm{cr}}$ indicates that the zero mode
 \begin{equation}
\psi_0(y)=\mathcal{N} f
\end{equation}
might be normalizable. Here normalization constant $\mathcal{N}$ is defined by
 \begin{eqnarray}
1&=&\int^{+\infty}_{-\infty}dy \psi_0^2(y)\nonumber\\
&=&\mathcal{N}^2 \int^{+\infty}_{-\infty}dx \mathcal{L}_Xe^{-A}\left(\frac{\partial_x\phi}{\partial_x A}\right)^2\nonumber\\
&=&-\frac4{\kappa}\mathcal{N}^2 \int^{+\infty}_{-\infty}dx e^{-A}\frac{\partial^2_x A}{(\partial_x A)^2},
\end{eqnarray}
where Eq.~\eqref{eqEin1} is used in the last step. Obviously, the normalization of the zero mode is determined only by the warp factor. After inserting the solution in Eq.~\eqref{SolWarp}, one immediately obtains the value of the normalization constant, which is, for example, $|\mathcal{N}|\approx 0.1669, 0.0998,0.0490$ for $\alpha=0, 0.2, 0.3$, respectively.  One can show that  as $\alpha$ approaches to $\alpha_{\rm{cr}}$, $|\mathcal N |$ approaches to zero.

\section{Conclusion}
\label{SecConc}
In this work, we studied self-gravitating kink solutions in a two-dimensional dilaton gravity model with K-field. We first conducted a general analysis on the linear stability of arbitrary static solutions under the dilaton gauge $\delta\varphi=0$. We found that the perturbation equation can be recast into a Schr\"odinger-like equation with factorizable Hamiltonian operator $\hat H =\hat{ \mathcal{A}}^\dagger\hat{ \mathcal{A}}$, provided the conditions in Eq.~\eqref{stabilityCond} are satisfied. Such a factorization ensures the non-negativity of the eigenvalues of the Hamiltonian, and therefore, the corresponding solutions are stable under linear perturbation.

As an explicit example, we studied an extended cuscuton model, whose Lagrangian contains both the cuscuton term and the standard kinetic term. We found that the dynamical equations of this model have a simple first-order formalism, from which analytic kink solutions can be easily constructed by choosing suitable superpotentials.

To analyze the impacts of the cuscuton term, we adopted the polynomial superpotential in Ref.~\cite{Zhong2021}. After comparing the present solution with the one of Ref.~\cite{Zhong2021}, we found that the cuscuton term can impact the asymptotic behavior of the warp factor, the shapes of both $V(\phi)$ and $V_{\text{eff}}$, and the normalization of the zero mode of the linear perturbation. However, we did not observe the creation of inner structure of the energy density within our parameter range.

It is also interesting to extend the present discussions to other 2D gravity models, for example, the 2D $f(R)$ gravity~\cite{Schmidt1999,NojiriOdintsov2020}. In contrast to the higher-dimensional case, one cannot find  an Einstein frame for the 2D $f(R)$ gravity by using conformal transformations. Thus, to discuss the linear stability of a solution, one has to confront with the higher-order frame.  Besides, 2D gravity models also set an ideal platform for the study of interaction between self-gravitating kinks. Similar problem has beed considered in 5D Einstein gravity, where curvature singularities are observed after the collision~\cite{TakamizuMaeda2006}. One may ask if these singularities are avoidable in some 2D models. These and other open problems are presently under consideration, and we hope to report on them in the near future.

\section*{Declaration of competing interest}
The authors declare that they have no known competing financial interests or personal relationships that could have appeared to influence the work reported in this paper.

\section*{Acknowledgments}

 This work was supported by the National Natural Science Foundation of China (Grant Nos.~12175169, 11847211, 11605127), Fundamental Research Funds for the Central Universities (Grant No.~xzy012019052), and China Postdoctoral Science Foundation (Grant No.~2016M592770).

\section*{Bibliography}



\end{document}